# Context Sensitivity without Contexts

## A Cut-Shortcut Approach to Fast and Precise Pointer Analysis

WENJIE MA[*] and SHENGYUAN YANG[*], Nanjing University, China

TIAN TAN[†], XIAOXING MA, CHANG XU, and YUE LI[†], Nanjing University, China

Over the past decades, context sensitivity has been considered as one of the most effective ideas for improving the precision of pointer analysis for Java. Different from the extremely fast context-insensitivity approach, context sensitivity requires every program method to be analyzed under different contexts for separating the static abstractions of different dynamic instantiations of the method's variables and heap objects, and thus reducing spurious object flows introduced by method calls. However, despite great precision benefits, as each method is equivalently cloned and analyzed under each context, context sensitivity brings heavy efficiency costs. Recently, numerous selective context-sensitive approaches have been put forth for scaling pointer analysis to large and complex Java programs by applying contexts only to the selected methods while analyzing the remaining ones context-insensitively; however, because the selective approaches do not fundamentally alter the primary methodology of context sensitivity (and do not thus remove its efficiency bottleneck), they produce much improved but still limited results.

In this work, we present a fundamentally different approach called CUT-SHORTCUT for fast and precise pointer analysis for Java. Its insight is simple: the main effect of cloning methods under different contexts is to filter spurious object flows that have been merged inside a callee method; from the view of a typical pointer flow graph (PFG), such effect can be simulated by cutting off (CUT) the edges that introduce precision loss to certain pointers and adding SHORTCUT edges directly from source pointers to the target ones circumventing the method on PFG. As a result, we can achieve the effect of context sensitivity without contexts. We identify three general program patterns and develop algorithms based on them to safely cut off and add shortcut edges on PFG, formalize them and formally prove the soundness. To comprehensively validate CUT-SHORTCUT's effectiveness, we implement two versions of CUT-SHORTCUT for two state-of-the-art pointer analysis frameworks for Java, one in Datalog for the declarative DOOP and the other in Java for the imperative TAI-E, and we consider all the large and complex programs used in recent literatures that meet the experimental requirements. The evaluation results are extremely promising: CUT-SHORTCUT is even able to run faster than context insensitivity for most evaluated programs while obtaining high precision that is comparable to context sensitivity (if scalable) in both frameworks. This is for the first time that we have been able to achieve such a good efficiency and precision trade-off for those hard-to-analyze programs, and we hope CUT-SHORTCUT could offer new perspectives for developing more effective pointer analysis for Java in the future.

---

[*]Both authors contributed equally to this paper.

[†]Corresponding author.

---

Authors' addresses: Wenjie Ma, 191250103@smail.nju.edu.cn; Shengyuan Yang, mf21320184@smail.nju.edu.cn, State Key Laboratory for Novel Software Technology, Nanjing University, China; Tian Tan, tiantan@nju.edu.cn; Xiaoxing Ma, xxm@nju.edu.cn; Chang Xu, changxu@nju.edu.cn; Yue Li, yueli@nju.edu.cn, State Key Laboratory for Novel Software Technology, Nanjing University, China.

---

## 1 INTRODUCTION

Pointer analysis is a family of static analysis techniques aimed at computing a set of abstract values that a program pointer may point to during program execution. The result of pointer analysis lays the groundwork for a wide range of static analysis applications such as bug detection [Cai et al. 2021; Chandra et al. 2009; Naik et al. 2006], security analysis [Arzt et al. 2014; Grech and Smaragdakis 2017], program optimization [Sridharan and Bodík 2006; Zhang et al. 2013], verification [Fink et al. 2008; Pradel et al. 2012], and understanding [Li et al. 2016; Sridharan et al. 2007]. Hence, it is not surprising that researchers have invested a substantial amount of effort in trying to build fast and precise algorithms for pointer analysis in the past decades [Smaragdakis and Balatsouras 2015].

Andersen-style pointer analysis [Andersen 1994] has been considered as the standard context-insensitive pointer analysis for Java [Smaragdakis and Balatsouras 2015; Sridharan et al. 2013]. We can see it as an iterative algorithm applied on a typical pointer flow graph (PFG) [Li et al. 2018a, 2020a; Tonella and Potrich 2005; WALA 2006] where nodes represent program pointers (including variables and instance fields) and edges represent subset constraints between pointers' points-to sets. During the analysis, abstracted objects are propagated along the edges of PFG. As the context-insensitivity approach does not distinguish between different call sites of the same method, incoming points-to sets are merged in callees; thus, it is extremely fast but with low precision.

For the sake of higher precision, context sensitivity is applied to pointer analysis. Basically, for each method call (say $c$) to a callee (say $m$), all elements in $m$ (like variables, instance fields and heap objects) will be cloned and analyzed under a context related to certain program element (say $e$) that distinguishes $c$ from other calls; accordingly, different forms of context sensitivity are presented like call-site sensitivity (if $e$ is a call site) [Sharir and Pnueli 1981; Shivers 1991], object sensitivity (if $e$ is an allocation site) [Milanova et al. 2002, 2005], and type sensitivity (if $e$ is a type) [Smaragdakis et al. 2011]. As the object flows merged in the callees are separated under different contexts, the spurious objects flows introduced by method calls can be filtered, thereby increasing the precision. However, context sensitivity comes with heavy efficiency costs as the analysis has to compute and maintain a huge number of context-sensitive intermediate results. As a result, context sensitivity, e.g., the most widely used 2obj (short for object sensitivity with context length being two), still fails to scale (cannot terminate in hours) for several complex programs in the standard DaCapo benchmarks [Jeong et al. 2017; Lhoták and Hendren 2006; Li et al. 2018b].

To scale for large and complex programs, in recent years, plenty of selective context-sensitivity approaches are proposed [Hassanshahi et al. 2017; Heo et al. 2017; Jeon et al. 2020; Jeong et al. 2017; Li et al. 2018a,b, 2020a; Lu et al. 2021; Oh et al. 2014, 2015; Smaragdakis et al. 2014; Tan et al. 2021; Wei and Ryder 2015]. Generally, they rely on a pre-analysis to select a set of methods that are critical to improving precision but do not jeopardize scalability when analyzed under context sensitivity. Then they only apply context sensitivity to those selected methods while analyzing the remaining ones context-insensitively. Although selective context sensitivity significantly improves the scalability of pointer analysis, its efficiency has not yet been fully unlocked as context sensitivity's core methodology has not been altered: we still have to spend potentially many computing resources replicating a set of methods and analyzing the program elements in each of those methods separately under different contexts, and once several of them threaten the scalability, the analysis runs the risk of being very slow or even not scalable [Li et al. 2020a; Smaragdakis et al. 2014].

In this work, we present a novel CUT-SHORTCUT approach to fundamentally change the status quo where pointer analysis for Java primarily relies on context sensitivity to acquire higher precision.

*Method.* The key observation is that imprecision occurs for context insensitivity when object flows are merged in a method $m$ and then the merged flows go outside $m$, introducing imprecision. Different from replicating a method $m$ into multiple copies to distinguish object flows from *different* call sites as context sensitivity does, our approach simulates context sensitivity's effect by *cutting* off the flows that introduce precision loss to certain pointers and adds the *shortcut* flows directly from exact source pointers to the target ones circumventing method $m$. As a result, we can achieve the effect of context sensitivity without applying contexts to $m$. To be more specific, CUT-SHORTCUT does not need a pre-analysis and it just performs the context-insensitive pointer analysis algorithm on a on-the-fly constructed graph $PFG'(N, E')$, which can be seen as a graph transformed from the original $PFG(N, E)$, where $N$ and $E$ denote the sets of nodes and edges of PFG respectively, and $E' = E \setminus \{edges\ to\ cut\ off\} \cup \{shortcut\ edges\ to\ add\}$ (in other words, for the standard context-insensitive analysis, only its PFG is changed to PFG'). Then we have two key questions to answer:

- *Which edges to cut off?* We cut off edges that may bring merged object flows inside a method to somewhere outside (e.g., to the left-hand side variable of the call site which receives the values of the return variables of the callee method). In other words, the mergence of object flows itself often does not harm precision as long as the chaos is kept inside the method. The main precision loss occurs when the merged object flows are propagated outside the method.
- *Where to add shortcuts?* We locate all the source nodes (say $s$) that precede the mergence of object flows inside the callee, and then create new edges between $s$ and the corresponding target nodes of the cut edges, for deriving sound and precise analysis results.

Although the basic idea seems simple, it is both challenging to identify *which edges to cut off* and *where to add shortcuts*. For the former, we need to identify edges that do great harm to precision and estimate whether cutting them off is beneficial, which takes a lot of maneuvering. For the latter, with some PFG edges cut off, if the added shortcuts can not capture all the necessary object flow sources, soundness will be harmed. To ensure soundness, instead of attempting to tackle all kinds of precision loss, in this work, we present three well-characterized program patterns that are suitable to be handled by our approach and accountable for a large portion of precision loss.

*Evaluation.* We prove CUT-SHORTCUT's soundness theoretically and further indirectly validate it by a recall experiment; moreover, in order to comprehensively validate its effectiveness, we implement two versions of CUT-SHORTCUT for two state-of-the-art Java pointer analysis frameworks, one in Datalog (≈100 Datalog rules) for the declarative Doop [Bravenboer and Smaragdakis 2009] and the other in Java (≈2000 lines of code) for the imperative TAI-E [Tan and Li 2022]. We consider all the large and complex Java programs used in recent literature that meet the experimental requirement, and compare CUT-SHORTCUT to context-insensitivity, conventional and selective context-sensitivity approaches. *The results are extremely promising: CUT-SHORTCUT is even able to run faster than context insensitivity for most evaluated programs while obtaining high precision that is comparable to context sensitivity (if scalable) in both frameworks.* This is for the first time that we have been able to achieve such a good efficiency and precision trade-off for those hard-to-analyze Java programs, and the experimental results can be obtained using the accompanying artifact [Ma et al. 2023].

## 2 MOTIVATING EXAMPLE

We use the example in Figure 1 to illustrate the basic idea of CUT-SHORTCUT and show how it differs from context insensitivity and context sensitivity for pointer analysis. Class Carton has a field item and provides setItem() and getItem() to modify and retrieve the item value respectively. In main(), two Item objects ($o_{16}, o_{21}$) are stored in the field item of two Carton objects ($o_{15}, o_{20}$) via method `setItem()` and retrieved later via `getItem()` (we use $o_i$ to denote the abstract heap object allocated at line $i$). After execution, `result1` (`result2`) will only point to $o_{16}$ ($o_{21}$). Below we explain how context insensitivity, sensitivity and our approach work for this example.

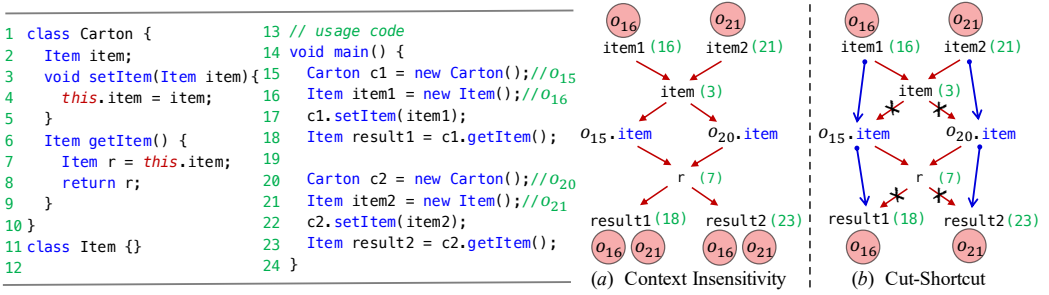

Fig. 1. An example to illustrate the basic idea of CUT-SHORTCUT.

*Context Insensitivity.* Figure 1(a) shows the simplified PFG for the left code (an edge $e$ of PFG means what pointed to by the source pointer of $e$ should flow to the target pointer of $e$). Since it does not distinguish the two calls of `setItem()`, $pt$(`item1`) (we use $pt(x)$ to denote $x$'s points-to set) and $pt$(`item2`) will merge into $pt$(`item`) (where `item` is the parameter at line 3), finally leading to imprecise points-to sets for $o_{15}$.`item` and $o_{20}$.`item` by the store statement at line 4. Moreover, two calls of `getItem()` are handled in the same manner, so that the merged objects ($\{o_{16}, o_{21}\}$) flow to $r$ by the load statement at line 7, rendering points-to sets of `result1` and `result2` imprecise.

*Context Sensitivity.* See Figure 1(a), the goal of context sensitivity is to separate the flows merged at `item` (line 3) and $r$ (line 7), and let $o_{16}$ only flows to `result1` and $o_{21}$ to `result2`. To distinguish the object flows from different call sites (namely, lines 17 and 22 for `setItem()`, and lines 18 and 23 for `getItem()`), every program element in `setItem()` (namely `item` and `this.item`) and `getItem()` (namely `this.item` and `r`) should be qualified with a context and analyzed separately, which is equivalently to analyze each of those methods twice under two different contexts. We can anticipate that the cost of computing and maintaining context-sensitive information would be very heavy in the real world, especially when the program is large or complex. Selective context sensitivity alleviates this problem but does not fundamentally change it, as even a small set of selected methods in a complex program may threaten scalability and blow up the analysis when they are analyzed context-sensitively [Li et al. 2020a; Smaragdakis et al. 2014].

*CUT-SHORTCUT.* Figure 1(b) depicts how CUT-SHORTCUT works, where the arrows with crosses represent the edges to cut off and the blue arrows indicate the shortcuts to add. The imprecision of points-to sets of $o_{15}$.`item` and $o_{20}$.`item` arises from edges `item` → $o_{15}$.`item` and `item` → $o_{20}$.`item` (by the store statement at line 4), so we cut them off and add shortcuts from the precise source (i.e., `item1` or `item2`) directly to the target node. Likewise, for nodes `result1` and `result2`, we cut off the edges starting from $r$ which introduces imprecision and add shortcuts $o_{15}$.`item` → `result1` and $o_{20}$.`item` → `result2`. Then on the modified PFG, by applying the same algorithm as context insensitivity (propagating the points-to results along the PFG edges), we will eventually derive points-to results as precise as context sensitivity for this example.

## 3 THE CUT-SHORTCUT APPROACH, INFORMALLY

We introduce the overall principle lying behind CUT-SHORTCUT (Section 3.1), on which the three program patterns, field access pattern (Section 3.2), container access pattern (Section 3.3) and local flow pattern (Section 3.4) are based.

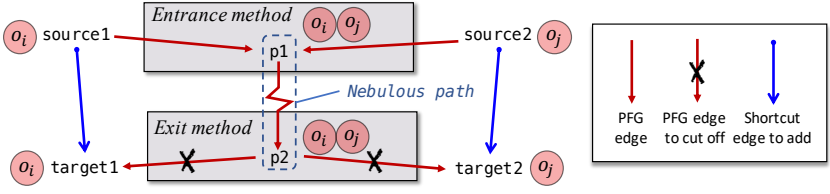

Fig. 2. The insight of Cut-Shortcut about which edges to cut off and where to add shortcut edges in PFG.

## 3.1 Overview

Briefly, Cut-Shortcut addresses the precision loss problem of context-insensitive pointer analysis by adhering to a general principle: cut the edges that carry merged flows and add shortcuts as substitution on PFG. To better unravel this principle, below we first clarify some terminologies.

*Definition 3.1 (Local/Non-local pointers).* If a pointer $p$ is method $m$'s local variable, then $p$ is local to $m$, otherwise $p$ is non-local to $m$, i.e., $p$ is a variable declared outside $m$ or is an instance field .

*Definition 3.2 (Nebulous path, Entrance and Exit).* If a path $p$ of PFG satisfies: (1) it starts with a confluence where two (or more) pointers non-local to *method m*, merge into a pointer $s$ local to $m$ (then $s$ is the start node of $p$), and (2) it ends with a divergence where a pointer $e$ local to *method n* (then $e$ is the end node of $p$) leads to two (or more) pointers non-local to $n$, then $p$ is a nebulous path, with $m$ being its Entrance and $n$ being its Exit.

Take nebulous path $\text{item} \rightarrow o_{15}.\text{item} \rightarrow \text{r}$ in Figure 1(a) as an example. $\text{item}$ and $\text{r}$ are the start and end nodes respectively of this path, and its Entrance is setItem() and Exit is getItem(). Path $\text{item}$ (at line 3) is also nebulous, whose start and end nodes are the same, namely $\text{item}$.

*Definition 3.3 (Target and Source pointers).* Given a nebulous path $p$, a Target pointer (Target for short) is any successor of $p$'s end node. In PFG, Target receives the merged points-to results from Exit. A Source pointer (Source for short) is any predecessor of the start node of $p$, whose points-to results flow to an Entrance and contribute to the merged flows on PFG.

Now we explain the insight of our principle in Figure 2. p1, p2, target1, target2, source1 and source2 are PFG nodes that represent program pointers (i.e., variables and instance fields) and $o_i$ and $o_j$ represent two abstract heap objects. Assume that p1 and p2 are the start and end nodes of a nebulous path, then source1 and source2 are recognized as Sources and target1 and target2 as Targets. During dynamic execution, target1 only points to $o_i$ that comes from source1; however, its points-to set encompasses both $o_i$ and $o_j$ in context insensitivity due to the nebulous path. In many cases of real programs, precise points-to results can be easily identified by human intuition with the comprehension of semantic information that are ignored by traditional pointer analysis. Our approach actually does the same thing by leveraging certain semantic clues to better capture the dynamic behavior of the program, so that we can effectively improve precision by cutting off the edges leaving the nebulous path (e.g., p2→target1/target2) and adding shortcuts to connect the precise Sources and Targets like source1→target1. Note that the operation "cutting off edges" is relative to the original PFG, and in Cut-Shortcut, we do not cut off edges that are already added to the PFG, since once an edge is added and then is removed later, it may introduce spurious object flows along such edges. Thus when we build PFG, we never add edges that should be cut off.

For better precision, we try to find as many Targets as possible and cut off the edges leading to them that bring precision loss, and for every Target, identify its corresponding Source(s) as accurately as possible and create shortcuts between Sources and Targets. However, this is hard as the real programs often exhibit convoluted behaviors and it is nearly impossible to offer a general scheme to handle all cases of precision loss [Li et al. 2018a, 2020a]; more importantly, we must ensure the analysis's soundness while manipulating PFG. Thus, we specify CUT-SHORTCUT for three program patterns, field access pattern, container access pattern and local flow pattern, which are commonly used in programs and result in a large portion of precision loss to pointer analysis. In the following subsections, we will discuss how to identify ENTRANCES and EXITS, SOURCES and TARGETS, and how to cut off edges and add shortcuts precisely and soundly for each pattern, respectively.

CUT-SHORTCUT is fundamentally different from selective context sensitivity [Jeon et al. 2019, 2020; Li et al. 2018a,b, 2020a; Smaragdakis et al. 2014; Tan et al. 2021]. Selective context-sensitive approaches select methods by different heuristics and apply contexts directly to them (equivalently to cloning the methods and analyzing them), and the main cost of these analyses is from cloning the methods and analyzing them multiple times. Differently, to avoid the overhead of cloning, CUT-SHORTCUT neither selects methods nor analyzes them context-sensitively; instead, it simulates the effect of context sensitivity by removing and adding edges of PFG (the challenge is to ensure soundness while removing edges), and propagates the points-to information on the modified PFG. No contexts are applied to any methods in CUT-SHORTCUT.

## 3.2 Field Access Pattern

There are two kinds of field accesses in Java, field store and field load (in this work, we focus on instance fields). Java programs often wrap field accesses in methods (e.g., setter/getter), so the client code needs to access the fields by calling these methods. However, in context insensitivity, object flows involved in the accesses to the fields of different objects, will be merged inside these methods, leading to imprecision. In this pattern, we tackle such precision loss under CUT-SHORTCUT's principle, as explained below. To ease understanding, we first introduce two basic cases about field stores (Section 3.2.1) and loads (Section 3.2.2) using the example of Figure 1 that relies on the information of field accesses and parameters, and describe a more general case in Figure 3 to show how our semantics-based scheme tracks the value flows from the caller (and the caller of the caller ...) where the values are finally stored in the target fields through parameters.

*3.2.1 Handling of Store.* We address the imprecision in the points-to results of instance fields. For a field store statement $s : x.f = y$, if objects pointed to by both $x$ and $y$ come from the parameters (including $this$ variable) of the method $m$ that contains $s$, then incoming object flows from arguments of different calls (of $m$) are merged inside $m$, which may cause imprecision in $pt(o_i.f)$ where $o_i \in pt(x)$. Line 4 in Figure 1 gives such an example: $this$ and $item$ are parameters of method setItem(). When analyzing setItem() context-insensitively, objects from call sites at lines 17 and 22 are merged in it, leading to imprecise result, i.e., $pt(o_{15}.item) = pt(o_{20}.item) = \{o_{16}, o_{21}\}$.

In CUT-SHORTCUT, our idea is to cut off the store edges which propagate the merged object flows to TARGETS, and find SOURCES at the call site of the method that contains the field store. We search for store statement $s : x.f = y$ that both $x$ and $y$ are parameters of $m$ (that contains $s$) and not redefined in $m$ (this guarantees that the values of both $x$ and $y$ all come from arguments of $m$'s call sites). If such $s$ is found, then different calls of $m$ form a nebulous path (with only one node $y$, e.g., the $item$ (3) in Figure 1(a)), making $m$ both ENTRANCE and EXIT. Accordingly, TARGETS are the instance fields being accessed by $x.f$ and stored at $s$ (e.g., $o_{15}.item$ and $o_{20}.item$ stored at line 4). To prevent merged object flows from being propagated to the TARGETS, we cut off the store edges to them (i.e., $item \rightarrow o_{15}.item$ and $item \rightarrow o_{20}.item$ in Figure 1).

Since the values of both $x$ and $y$ come from arguments of $m$'s call sites, we can find the matched TARGETS and SOURCE at each call site. Namely, the SOURCE is the argument passed to $y$ (e.g., the argument is the item1 at line 17 and $y$ is the parameter $item$ at line 3) and the TARGET is $o_i.f$, where $o_i \in pt(a)$ and $a$ is the argument passed to $x$ (e.g., the argument $a$ is c1 at line 17 and $x$ is the parameter this). For example, at line 17, item1 is the SOURCE for TARGET $o_{15}$.item (as $pt(c1) = \{o_{15}\}$), and likewise, item2 is the SOURCE for TARGET $o_{20}$.item at line 22 (as $pt(c2) = \{o_{20}\}$). Thus, we add shortcut edges item1 $\rightarrow o_{15}$.item and item2 $\rightarrow o_{20}$.item, and achieve precise result, i.e., $pt(o_{15}.\text{item}) = \{o_{16}\}$ and $pt(o_{20}.\text{item}) = \{o_{21}\}$ (depicted at the top half of Figure 1(b)).

### 3.2.2 Handling of Load.

We address the imprecision in the points-to results for LHS variables whose values are returned from the methods that load instance fields. For a field load statement $s : x = y.f$, if objects pointed to by $y$ come from a parameter (including this variable) of the method $m$ that contains $s$, and $x$ is the return variable of $m$, then the objects loaded from $y.f$ (where $y$ points to the incoming objects from arguments of different calls of $m$) are merged inside $m$, which may cause imprecision in points-to sets for the LHS variables of $m$'s call sites. Line 7 in Figure 1 gives such an example: this is a parameter of method getItem() and r is its return variable. When analyzing getItem() context-insensitively, objects from call sites at lines 18 and 23 (i.e., $o_{15}$ and $o_{20}$) are merged in it, and objects loaded from $o_{15}$ and $o_{20}$ (i.e., $o_{16}$ and $o_{21}$) are also merged at r and propagated to result1 and result2, leading to imprecise result $pt(\text{result1}) = pt(\text{result2}) = \{o_{16}, o_{21}\}$.

In CUT-SHORTCUT, our idea is to cut off the return edges from $m$ (that return values loaded from fields) which propagate merged object flows to TARGETS, and find SOURCES at $m$'s call sites. We search for load statement $s : x = y.f$ that the values of both $x$ and $y$ are directly related to the variables of the call sites of $m$ (that contains $s$), i.e., $y$ is a parameter of $m$ and not redefined in $m$ (this guarantees that the value of $y$ all come from arguments of $m$'s call sites), and $x$ is the return variable of $m$. If such $s$ and $m$ are found, we cut off the return edges from $m$ to LHS variable (i.e., the TARGET of each $m$'s call site to avoid propagation of merged flows, and then identify the SOURCES at the call site, i.e., $o_i.f$ where $o_i \in pt(a)$ and $a$ is the argument passed to $y$ (e.g., $a$ is c1 at line 18 and $y$ is this). For example, for field load at line 7 in Figure 1, we cut off edges r $\rightarrow$ result1 and r $\rightarrow$ result2. For result1, we find its SOURCE, $o_{15}$.item, at line 18 (as $pt(c1) = \{o_{15}\}$), and add shortcut edge $o_{15}$.item $\rightarrow$ result1. Similarly, we add $o_{20}$.item $\rightarrow$ result2 at line 23. Finally, we obtain precise result, $pt(\text{result1}) = \{o_{16}\}$ and $pt(\text{result2}) = \{o_{21}\}$ (see bottom half of Figure 1(b)).

### 3.2.3 Handling of Nested Call for Field Access.

In real-world programs, field access becomes complicated when nested method calls are involved. For instance, when a constructor is called, it may call another constructor or setter method where the store statement is finally executed. In addition, when inheritance is used, the usage of nested calls for field access becomes common.

```
1 class A {              6 // usage code
2   T f;                 7 T t1 = new T(); //o₇
3   A(T t){this.set(t);} 8 A a1 = new A(t1); //o₈
4   set(T p){this.f=p;}  9 T t2 = new T(); //o₉
5 }                      10 A a2 = new A(t2); //o₁₀
```

Fig. 3. An example of nested calls for field access.

For example, in Figure 3, A.set() is a setter called by A's constructor (line 3). There are two nested calls to it, 8 $\rightarrow$ 3 $\rightarrow$ A.set() and 10 $\rightarrow$ 3 $\rightarrow$ A.set() (a call site is represented by its line number). Here, if we add shortcut edges, t $\rightarrow o_8$.f and t $\rightarrow o_{10}$.f, at the direct call site of A.set() (line 3), we still lose precision as both this and t are parameters of A(), and object flows from call sites at lines 8 and 10 are merged here, leading to imprecise result $pt(o_8.f) = pt(o_{10}.f) = \{o_7, o_9\}$. Similarly, nested call for field load may also cause precision loss. To handle nested call for field access, we extend our approach to a more general manner for better precision. Specifically, for a method that contains field access, we analyze its nested calls (from the method itself to its caller, caller's caller, and so on, if possible) to cut off edges and add shortcut edges at proper call sites. For the field store in A.set(), we trace to the call sites at lines 8 and 10, add shortcut edges t1 $\rightarrow o_8$.f and t2 $\rightarrow o_{10}$.f, and derive precise result $pt(o_8.f) = \{o_7\}$ and $pt(o_{10}.f) = \{o_9\}$. For details of our approach to handle nested call for field store and load, please refer to Section 4.2.

## 3.3 Container Access Pattern

Containers (e.g., `ArrayList`) are pervasively used in Java programs. As numerous objects flow in and out of containers, they are crucial to the precision of pointer analysis. In context insensitivity, objects stored in different containers may flow to and merge in the same container methods, and horrendous precision loss may be introduced when the merged objects flow out. In order to separate object flows in different containers, conventional approaches improve precision by analyzing container methods context-sensitively. However, as pointed out by recent work [Antoniadis et al. 2020; Fegade and Wimmer 2020], applying context sensitivity to the container methods will bring heavy efficiency costs, especially for large programs. As a result, to analyze containers precisely while reducing analysis overhead, researchers [Antoniadis et al. 2020; Fegade and Wimmer 2020] propose to manually rewrite container implementations for composing code models that are simpler than the original implementations while still preserving the side effects of containers to pointer analysis. Then, they apply context sensitivity to the rewritten container code during the analysis. In Cut-Shortcut, we also need to handle containers for obtaining good precision. But we will do it differently by adopting a scheme under the principle of Cut-Shortcut.

### 3.3.1 Handling of Containers.

Figure 4 shows an `ArrayList` example of container (just focus on lines 1-9 for now). We create two `ArrayList`s, $o_1$ (line 1) and $o_6$ (line 6), and add two objects $o_2$ (line 3) and $o_6$ (line 8) to them, respectively. At lines 4 and 9, we retrieve the elements from $o_1$ and $o_6$ and assign them to x and y separately. Clearly, x (y) only points to $o_2$ ($o_7$) at run time.

Container implementations typically provide APIs to add elements to, or retrieve elements from containers. Thus it is natural to consider the addition and retrieval APIs as the Entrance and Exit methods for container respectively, since objects are stored into and flow out of containers via them. We denote the call sites of Entrance and Exit as Entrance$_{cs}$ and Exit$_{cs}$, respectively; for example, as commented in Figure 4, add() (get()) is the Entrance (Exit) of `ArrayList`.

```
1  ArrayList l1 = new ArrayList(); //o₁,ptₕ(l1)={o₁}
2  Object a = new Object(); //o₂
3  l1.add(a); //ENTRANCE, a is SOURCE
4  Object x = l1.get(0); //EXIT, x is TARGET
5
6  ArrayList l2 = new ArrayList(); //o₆,ptₕ(l2)={o₆}
7  Object b = new Object(); //o₇
8  l2.add(b); //ENTRANCE, b is SOURCE
9  Object y = l2.get(0); //EXIT, y is TARGET
10
11 Iterator it1 = l1.iterator(); //ptₕ(it1)={o₁}
12 Object r1 = it1.next(); //EXIT, r1 is TARGET
13 Iterator it2 = l2.iterator(); //ptₕ(it2)={o₆}
14 Object r2 = it2.next(); //EXIT, r2 is TARGET
```

Fig. 4. An example of `ArrayList`.

Then, the related argument of a Entrance$_{cs}$ is treated as a Source pointer (say Entrance$_{cs}^{arg}$); for example, a at line 3 is the Entrance$_{cs}^{arg}$ for the call site of add() at line 3. The LHS variable of a Exit$_{cs}$ is treated as a Target pointer (say Exit$_{cs}^{lhs}$); for example, x at line 4 is the Exit$_{cs}^{lhs}$ for the call site of get() at line 4. In Cut-Shortcut, we cut off the return edges from Exits to the Exit$_{cs}^{lhs}$ so that the merged object flows inside a container will not be propagated to the Exit$_{cs}^{lhs}$. Then we add a shortcut edge from each Entrance$_{cs}^{arg}$ to the corresponding Exit$_{cs}^{lhs}$ for complementing the points-to sets of Exit$_{cs}^{lhs}$.

In context-insensitivity, $o_2$ and $o_7$ will be merged in `ArrayList`'s methods, and both flow to x and y via get(), leading to imprecision. In Cut-Shortcut, to avoid precision loss, we cut off edges returned from get() to both Exit$_{cs}^{lhs}$, x and y. To add shortcut edges, we need to find out the matched Entrance$_{cs}^{arg}$ and Exit$_{cs}^{lhs}$ pair (if they are related to the same container, they are matched). In our case, a and x is a matched Entrance$_{cs}^{arg}$ and Exit$_{cs}^{lhs}$ pair, as their corresponding Entrance$_{cs}$ (add() at line 3) and Exit$_{cs}$ (get() at line 4) are related to the same container ($o_1$).

A straightforward approach to determine whether Entrance$_{cs}$ and Exit$_{cs}$ are related is to examine whether the points-to sets of the corresponding Entrance$_{cs}^{rv}$ and Exit$_{cs}^{rv}$ overlap, where Entrance$_{cs}^{rv}$ and Exit$_{cs}^{rv}$ are the receiver variables of Entrance$_{cs}$ (e.g., l1 of add() at line 3) and $\textsc{Exit}_{cs}$ (e.g., $\text{l2}$ of $\text{get()}$ at line 9), respectively. However, this approach cannot handle the other container cases like iterators and $\text{Map}$'s $\text{keySet()}$, etc. Thus, we introduce a concept called *pointer-host map* ($pt_H$) to address the above issue uniformly.

$pt_H$ maps a host-related pointer to a set of container objects (called host) related to it. For example, $\text{ArrayList}$ $o_1$ is assigned to $\text{l1}$ (line 1), and thus we have $pt_H(\text{l1}) = \{o_1\}$ (we will introduce more mapping rules later). Then if $pt_H$ ($\textsc{Entrance}_{cs}^{rv}$) and $pt_H$ ($\textsc{Exit}_{cs}^{rv}$) overlap (namely, they may share the same host container object), we add a shortcut edge from the corresponding $\textsc{Entrance}_{cs}^{arg}$ to $\textsc{Exit}_{cs}^{lhs}$. In the above example, as $pt_H(\text{l1})$ (where $\text{l1}$ is the receiver variable of $\text{add()}$ at line 3) and $pt_H(\text{l1})$ (where $\text{l1}$ is the receiver variable of $\text{get()}$ at line 4) overlap, we add a shortcut edge $\text{a} \rightarrow \text{x}$; similarly for $\text{b} \rightarrow \text{y}$. As a result, $\text{x}$ ($\text{y}$) can precisely point to only $o_2$ ($o_7$). In this case, $pt_H$ serves the same functionality as typical points-to relation, $pt$. We will explain the other usages of $pt_H$ below.

### 3.3.2 Handling of Host-Dependent Objects.

Lines 11-14 in Figure 4 show a very common usage of container, i.e., accessing elements via iterators. To comprehensively handle containers for better precision, we need to take such usage into account. Typically, each iterator depends on a container (i.e., the host), and the elements retrieved from the iterator are the ones stored in the host. We call the iterator objects the *host-dependent objects*[1].

To handle host-dependent objects, our principle remains the same as the one in Section 3.3.1: we cut off the return edges from $\textsc{Exits}$ to the $\textsc{Exit}_{cs}^{lhs}$ and add a shortcut edge from each $\textsc{Entrance}_{cs}^{arg}$ to the corresponding $\textsc{Exit}_{cs}^{lhs}$ if $pt_H$ ($\textsc{Entrance}_{cs}^{rv}$) and $pt_H$ ($\textsc{Exit}_{cs}^{rv}$) overlap. The difference is that we need to expand the set of $\textsc{Entrance}$, $\textsc{Exit}$ and $pt_H$ for host-dependent objects. In our example, container elements can be retrieved by $\text{next()}$ (lines 12 and 14), thus method $\text{next()}$ is added as an $\textsc{Exit}$ (no expansion to $\textsc{Entrance}$ for this case). As for $pt_H$, in our case, we need to expand this mapping relation by (1) adding a host-related pointer, say $Hptr$ (e.g., the iterator pointer $\text{it1}$ at line 11) to its key set, and (2) adding its related host object, say $H$ (e.g., the $\text{ArrayList}$ object $o_1$) to its value set. Then the last question is how to identify $H$ via $Hptr$? To answer this question, we first define a notation called $\textsc{Transfer}$ methods. If a host-dependent object (the object pointed to by $Hptr$) is created by invoking a method $m$ of the class of host $H$, we call $m$ a $\textsc{Transfer}$. In our example, $\text{iterator()}$ at line 11 is a $\textsc{Transfer}$ as it is a method of $H$'s class ($\text{ArrayList}$) and invoking it creates/returns a $\text{iterator}$ object that is pointed to by $Hptr$ ($\text{it1}$ at line 11). So to map $Hptr$ to $H$, for each call site of $\textsc{Transfer}$, we propagate the host objects of its receiver variable ($pt_H(\text{l1})$ at line 11) to its LHS variable ($\text{it1}$ at line 11). As a result, we have new $pt_H$ relation $pt_H(\text{it1})$ $= \{o_1\}$ as $pt_H(\text{it1}) \supseteq pt_H(\text{l1}) = \{o_1\}$. We call $\text{iterator()}$ a $\textsc{Transfer}$ as it transfers the host objects from its receiver variable to its LHS variable. Note that building $pt_H$ requires the results of pointer analysis, and our detection of container access pattern needs to be solved on the fly with pointer analysis. By leveraging pointer analysis, our approach can also handle the cases involving aliasing when building $pt_H$ and its details will be formalized by rules in Section 4.3.

In our scheme, we only need to specify which APIs are $\textsc{Entrances}$, $\textsc{Exits}$ or $\textsc{Transfers}$, and our analysis will take care of the rest automatically. For generality, we only consider the JDK containers, the APIs of which are stable. As they are also well documented and it is straightforward to identify the types of relevant APIs by their names, it only took one of the authors five hours totally to specify the needed APIs in JDK. Compared to past work [Antoniadis et al. 2020; Fegade and Wimmer 2020] that requires to rewrite code implementations for related container APIs, our scheme is more general and simpler.

---

[1]Actually, host-dependent objects include not only iterators but also some other objects that depend on certain container, such as the collection views of $\text{Map}$ (e.g., the return value of $\text{Map.keySet()}$), which can also be handled by our approach.

### 3.4 Local Flow Pattern

This is a simple pattern to address the precision loss case where the values of a method's parameters flow to the return variable via a series of local assignments. When such a method is called from multiple call sites, their argument values are merged into the method, and propagated together to the LHS variable at each call site, leading to imprecise points-to results. We use an example in Figure 5 to illustrate this pattern. In method select(), the values of its parameters p1 and p2 will flow to return variable r. It has two call sites at lines 12 and 16. In context insensitivity, objects $o_{10}, o_{11}, o_{14}$ and $o_{15}$ will be passed to and merged in select(), and then flow together to both r1 and r2 imprecisely, as r1 (r2) only points to $o_{10}$ or $o_{11}$ ($o_{14}$ or $o_{15}$) in any execution.

```
1 A select(A p1,A p2){   9 // usage code
2   A r;                 10 A a1 = new A(); //o10
3   if (…)               11 A a2 = new A(); //o11
4       r = p1;          12 A r1 = select(a1, a2);
5   else                 13
6       r = p2;          14 A a3 = new A(); //o14
7   return r;            15 A a4 = new A(); //o15
8 }                      16 A r2 = select(a3, a4);
```

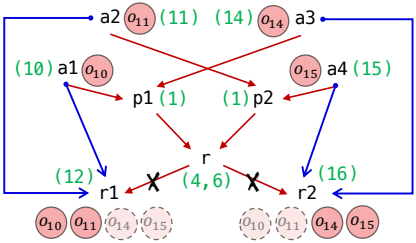

Fig. 5. An example of local flow pattern.

In Cut-Shortcut, for this pattern, Entrance and Exit are the same method, whose parameter values flow to the return variable. We focus on the cases that the return values of a method must all come from its parameters via local assignments (excluding other sources of return values such as field loads and method calls), and the detection of such cases relies on an intraprocedural value flow analysis. Hence, we can safely use arguments (that correspond to the parameters) as the Sources, and add shortcut edges from them to the Targets, i.e., the LHS variables of the call sites. In this example, we cut off the return edges from r, and add shortcut edges a1 → r1 and a2 → r1 (a3 → r2 and a4 → r2) for the call site at line 12 (16). As a result, r1 (r2) points to only $\{o_{10}, o_{11}\}$ ($\{o_{14}, o_{15}\}$) in Cut-Shortcut, which is precise. We formally describe how to analyze this pattern in Section 4.4.

*Limitations.* Note that our approach is only for Java, thus it cannot be directly applied to other programming languages; however, the high-level idea of Cut-Shortcut (Figure 2), namely removing imprecision loss edges and adding semantics-preserving edges to the PFG might inspire techniques for improving the precision of pointer analysis for other languages. Regarding the applicability of Cut-Shortcut to other abstractions like context sensitivity, our current approach is not designed to be pluggable into other (selective) context-sensitive analyses. Rather than selecting methods and analyzing them context-sensitively, we directly manipulate PFG edges without applying any contexts. But it would be interesting to try such a combination. For example, for the methods whose PFG edges are not affected by our approach, but considered by other selective context-sensitivity approaches [Jeon et al. 2019; Li et al. 2018a,b, 2020a; Smaragdakis et al. 2014], we can analyze them context-sensitively for possibly better precision. As for flow sensitivity, like almost all whole-program pointer analyses for Java in past literature [Jeon and Oh 2022; Lu et al. 2021; Smaragdakis and Balatsouras 2015; Sridharan et al. 2013; Tan et al. 2021], our approach is also flow-insensitive. It is non-trivial to apply our current approach to flow-sensitive analysis, but it would be interesting to try it in the future.

## 4 FORMALISM AND SOUNDNESS

In this section, we formalize pointer analysis with our Cut-Shortcut approach (in Section 4.1) and the three patterns (in Sections 4.2– 4.4), and prove the soundness of Cut-Shortcut (in Section 4.5).

| methods | $m \in \mathbb{M}$ | $i \xrightarrow{call} m$ | a call edge from call site $i$ to method $m$ |
| instruction labels | $i, j \in \mathbb{L}$ | $i_{ak}$ | the $k$-th argument of call site $i$ |
| variables | $x, y \in \mathbb{V}$ | $m_{pk}$ | the $k$-th formal parameter of method $m$ |
| heap objects | $o_i, o_j \in \mathbb{O}$ | $m_{ret}$ | the return variable of method $m$ |
| fields | $f \in \mathbb{F}$ | $def_x$ | the set of statements that define variable $x$ |
| pointers | $x, o_i.f \in \mathbb{P} = \mathbb{V} \cup (\mathbb{O} \times \mathbb{F})$ | PFG | $G = (N, E)$ |
| points-to relations | $pt : \mathbb{P} \to \mathcal{P}(\mathbb{O})$ | nodes | $n \in N = \mathbb{P}$ |
| | | edges | $s \to t \in E \subseteq N \times N$ |

Fig. 6. Domains and notations used in formalism.

$$\frac{i : x = new\, T()}{o_i \in pt(x)} \text{ [New]} \qquad \frac{x = y}{y \to x \in E} \text{ [Assign]} \qquad \frac{s \to t \in E}{pt(s) \subseteq pt(t)} \text{ [Propagate]} \qquad \frac{e \in E_{SC}}{e \in E} \text{ [Shortcut]}$$

$$\frac{x = y.f \quad o_i \in pt(y)}{o_i.f \to x \in E} \text{ [Load]} \qquad \frac{i : x.f = y \quad o_j \in pt(x) \quad \boxed{i \notin cutStores}}{y \to o_j.f \in E} \text{ [Store]}$$

$$\frac{i : x = y.m(arg_1, ..., arg_n) \quad o_j \in pt(y) \quad m' = dispatch(o_j, m)}{i \xrightarrow{call} m' \quad o_j \in pt(m'_{p0})} \text{ [Call]}$$

$$\frac{i : x = y.m(arg_1, ..., arg_n) \quad i \xrightarrow{call} m'}{\forall\, 1 \le k \le n : arg_k \to m'_{pk} \in E} \text{ [Param]} \qquad \frac{i : x = y.m(...) \quad i \xrightarrow{call} m' \quad \boxed{m'_{ret} \notin cutReturns}}{m'_{ret} \to x \in E} \text{ [Return]}$$

Fig. 7. Rules of pointer analysis, with Cut-Shortcut.

## 4.1 Pointer Analysis with Cut-Shortcut

In this section, we formalize pointer analysis integrated with our Cut-Shortcut approach. For illustrative purpose, we give in Figure 6 the domains and notations we will use in our formalism, which are mostly self-explanatory. Here, $pt(x)$ gives the analysis result: it maps each pointer $x$ to its points-to set ($\mathcal{P}(\mathbb{O})$ denotes power set of $\mathbb{O}$), and for $m_{pk}$ and $i_{ak}$, we use $m_{p0}$ ($k$=0) to denote this variable of method $m$, and $i_{a0}$ to denote receiver variable of call site $i$.

*Pointer Analysis, in a PFG View.* Figure 7 shows our formalism of pointer analysis (for now, just ignore rule [Shortcut] and premises in black boxes which are related to Cut-Shortcut). Essentially, our formalism is equivalent to the ones appearing in existing literature [Milanova et al. 2005; Smaragdakis and Balatsouras 2015; Sridharan et al. 2013] as they all express Andersen-style pointer analysis for Java. However, to integrate with Cut-Shortcut, we explicitly express the *subset constraints* among pointers by the edges of the PFG (pointer flow graph). Specifically, all nodes of PFG are pointers, and by [Propagate], an edge $s \to t$ in the PFG indicates that the objects pointed to by $s$ should also be pointed to by $t$ (i.e., a subset constraint). Other rules describe how to generate PFG edges for different kinds of statements: object allocation ([New]), local assignment ([Assign]), instance field load ([Load]) and store ([Store]), and method invocation ([Call],[Param], [Return]).

*Cut-Shortcut.* Cut-Shortcut is a general approach with one principle throughout: first cut off the edges that may introduce precision loss, and then add shortcut edges to connect proper Sources to Targets. To formalize this principle, we define three sets, *cutStores* and *cutReturns* for cutting off edges, and $E_{SC}$ for adding shortcut edges, as explained below.

• *cutStores* is a set of store statements (identified by their labels). By [Store], if a store statement $j$ is in *cutStores*, then the store edges that should have been generated for $j$ will be cut off.

$$\frac{i \xrightarrow{call} m \quad m_{pk} = x \quad def_x = \varnothing}{i_{ak} \mapsto x} \text{ [Arg2Var]} \qquad \frac{i : x.f = y \quad j_{ak_1} \mapsto x \quad j_{ak_2} \mapsto y}{i \in cutStores \quad \langle j_{ak_1}, f, j_{ak_2} \rangle \in tempStores} \text{ [CutStore]}$$

$$\frac{\langle base, f, from \rangle \in tempStores \quad j_{ak_1} \mapsto base \quad j_{ak_2} \mapsto from}{\langle j_{ak_1}, f, j_{ak_2} \rangle \in tempStores} \text{ [PropStore]}$$

$$\frac{\langle base, f, from \rangle \in tempStores \quad base \text{ in } m \quad o_j \in pt(base)}{((def_{base} \neq \varnothing) \vee (def_{from} \neq \varnothing) \vee (\nexists k : m_{pk} = base) \vee (\nexists k : m_{pk} = from))}{from \to o_j.f \in E_{SC}} \text{ [ShortcutStore]}$$

Fig. 8. Rules for handling field stores in field access pattern.

$$\frac{((i : to = base.f) \vee (\langle to, base, f \rangle \in tempLoads))}{m_{ret} = to \quad j : r = \_ \quad j_{ak} \mapsto base \quad o_n \in pt(base)}{m_{ret} \in cutReturns \quad \langle r, j_{ak}, f \rangle \in tempLoads \quad o_n.f \to to \in returnLoadEdges} \text{ [CutPropLoad]}$$

$$\frac{\langle to, base, f \rangle \in tempLoads}{o_i \in pt(base)}{o_i.f \to to \in E_{SC}} \text{ [ShortcutLoad]} \qquad \frac{i : r = \_ \quad i \xrightarrow{call} m \quad m_{ret} \in cutReturns}{n \to m_{ret} \in (E \setminus returnLoadEdges)}{n \to r \in E_{SC}} \text{ [RelayEdge]}$$

Fig. 9. Rules for handing field loads in field access pattern.

- *cutReturns* is a set of return variables. By [Return], if a return variable $m_{ret}$ is in *cutReturns*, then the return edges from $m_{ret}$ to the LHS variables of $m$'s call sites will be cut off.
- $E_{SC}$ is a set of shortcut edges. By [Shortcut], if an edge $s \to t$ is added to $E_{SC}$, it will also be added to the PFG, and propagate the objects pointed to by to $s$ to $t$.

Next, we describe how these three sets are computed for the three patterns in Sections 4.2–4.4.

## 4.2 Field Access Pattern

In this section, we formalize our handling of store and load, including nested call for field access.

*4.2.1 Handling of Store.* Figure 8 gives the rules for computing *cutStores* and $E_{SC}$ to handle field stores. Intuitively, we cut off the PFG edges introduced by a store statement and add proper shortcut edges at the call site (of the method containing the store) when we can confirm that the base and RHS variables of the store (e.g., $x$ and $y$ of $x.f = y$) always point to the same objects of the arguments of the call site at each invocation. To help establish shortcut edges, we define *tempStores*, which is a set of triples like $\langle base, f, from \rangle$ that represents a (potentially) store operation $base.f = from$. To handle nested calls (mentioned in Section 3.2.3) for better precision, we construct and propagate temp stores along the call chains, from the innermost callee (containing the actual store statement) to the outermost callers, as explained below.

*Cut.* To simplify rules, we introduce notation $\mapsto$ defined by [Arg2Var], where $i_{ak} \mapsto x$ states that the values of $k$-th argument of call site $i$ flow to variable $x$ of method $m$, and $x$ is not redefined in $m$, i.e., $x$ points to the same object as $i_{ak}$ at each invocation of $m$. Now we define *cutStores* in [CutStore]. For a store statement $i : x.f = y$, if the values of both $x$ and $y$ come from the arguments of call site of the method containing $i$, then we add $i$ to *cutStores* to cut off the PFG edges introduced by $i$. Besides, [CutStore] generates temp stores which may be further propagated to the callers.

*Shortcut.* We define two rules [PropStore] and [ShortcutStore] to derive $E_{SC}$ for field stores. [PropStore] describes the process of constructing new temp stores and propagating them from callee to caller recursively. In [ShortcutStore], if the disjunction holds, it means that the temp store cannot be further propagated to the callers, then we generate shortcut edges for the temp store.

$$\frac{\langle m, \_ \rangle \in \textsc{Exits}}{m_{ret} \in cutReturns} \; [\textsc{CutContainer}]$$

$$\frac{h \overset{c}{\Leftarrow} s \quad h \overset{c}{\Rightarrow} t}{s \to t \in E_{SC}} \; [\textsc{ShortcutContainer}]$$

$$\frac{i : x = y.m(...) \quad i \overset{call}{\longrightarrow} m'}{\langle m', k, c \rangle \in \textsc{Entrances} \quad h \in pt_H(y)}{h \overset{c}{\Leftarrow} i_{ak}} \; [\textsc{HostSource}]$$

$$\frac{i : x = y.m(...) \quad i \overset{call}{\longrightarrow} m'}{\langle m', c \rangle \in \textsc{Exits} \quad h \in pt_H(y)}{h \overset{c}{\Rightarrow} x} \; [\textsc{HostTarget}]$$

$$\frac{o_i \in pt(x) \quad typeof(o_i) <: \texttt{Collection}}{o_i \in pt_H(x)} \; [\textsc{ColHost}]$$

$$\frac{o_i \in pt(x) \quad typeof(o_i) <: \texttt{Map}}{o_i \in pt_H(x)} \; [\textsc{MapHost}]$$

$$\frac{i : x = y.m(...) \quad i \overset{call}{\longrightarrow} m'}{m' \in \textsc{Transfers} \quad h \in pt_H(y)}{h \in pt_H(x)} \; [\textsc{TransferHost}]$$

$$\frac{s \to t \in E \quad \neg(i : t = y.m(...) \land}{i \overset{call}{\longrightarrow} m' \land m' \in \textsc{Transfers} \land s = m'_{ret})}{pt_H(s) \subseteq pt_H(t)} \; [\textsc{PropHost}]$$

Fig. 10. Rules for container access pattern.

*4.2.2 Handling of Load.* Figure 9 shows the rules for handling field loads. Specifically, if return values of a method $m$ are loaded from the objects that are passed to $m$ (as arguments), then they may cause precision loss as arguments from different call sites of $m$ are merged inside $m$. For such cases, we cut off the PFG edges from return variables of $m$ to LHS variables of its call sites by adding the return variables to *cutReturns*. Similar to handling of stores, we define *tempLoads* to help adding shortcuts, which is a set of triples like $\langle to, base, f \rangle$ that represents a load operation $to = base.f$. Also, to handle nested calls (mentioned in Section 3.2.3), we generate and propagate temp loads along the call chains.

*Cut.* In handling of load, the rules for cutting off edges and propagating temp loads are almost identical, hence we merge them into one rule [CutPropLoad] to avoid redundancy. Based on the insight explained above, [CutPropLoad] cuts off the return edges from the return variable $m_{ret}$ to the LHS variables of call sites of $m$, and constructs and propagates temp loads from callee to caller recursively, when possible. The rightmost premise and conclusion in [CutPropLoad] record the load edges to the return variable ($to$) in a set *returnLoadEdges*, which will be used to ensure soundness as discussed at the end of this section.

*Shortcut.* In [ShortcutLoad], we simply generate shortcut edges for triples in *tempLoads*.

In [CutPropLoad], we cut off all return values of a method $m$ to the LHS variables of its call sites, and do not require that $m_{ret}$ can only be defined by the load statement (such restriction would noticeably weaken precision improvements). However, in some cases, not all of the return values come from field loads (e.g., $m$ contains both $x = y.f$ and $x = z$ where $x$ is $m_{ret}$). Hence, we define [RelayEdge] to ensure soundness. If the values of a return variable $m_{ret}$ come from the pointers irrelevant to the field loads (i.e., not source pointers of edges in *returnLoadEdges*), then we add shortcut edges to connect the pointers to the LHS variable of $m$'s call sites.

## 4.3 Container Access Pattern

To formalize container access pattern, we classify container elements into three categories (denoted by $c$): values in a collection, keys in a map, and values in a map. The category is especially useful for maps as it can tell the analysis whether an Entrance/Exit method manipulates keys or values of a map. We define three input relations introduced in Section 3.3.

- Entrances, a set of triples like $\langle m, k, c \rangle$, which means that $m$ is a container Entrance method, and the values passed to its $k$-th parameter are of category $c$.
- Exits, a set of pairs like $\langle m, c \rangle$, which means that $m$ is a container Exit method that returns values of category $c$.

$$\frac{def_{m_{pk}} = \varnothing}{\langle m, k \rangle \rightarrowtail m_{pk}} \text{ [Param2Var]} \qquad \frac{\forall i \in def_x : (i : x = y) \wedge (\langle m, \_ \rangle \rightarrowtail y)}{\langle m, k \rangle \rightarrowtail x} \text{ [Param2VarRec]}$$

$$\frac{\langle m, \_ \rangle \rightarrowtail m_{ret}}{m_{ret} \in cutReturns} \text{ [CutLFlow]} \qquad \frac{i \xrightarrow{call} m \quad i : r = \_ \quad \langle m, k \rangle \rightarrowtail m_{ret}}{i_{ak} \rightarrow r \in E_{SC}} \text{ [ShortcutLFlow]}$$

Fig. 11. Rules for local flow pattern.

- TRANSFERS, a set of TRANSFER methods that transfer hosts from receiver variable of the call site to its LHS variable, e.g., from $y$ to $x$ for $x = y.m(...)$.

Now, we can formalize container access pattern as shown in Figure 10.

*Cut.* In [CutContainer], we simply cut all return edges from the methods specified in EXITs.

*Shortcut.* In [ShortcutContainer], we add shortcut edges based on two relations as premises, denoted by $h \overset{c}{\Leftarrow} s$ and $h \overset{c}{\Rightarrow} t$, which associate hosts (containers) to their SOURCES and TARGETS. The former states that $s$ is a SOURCE of host $h$, i.e., $s$ points to the objects flowing into $h$ and the objects become elements of $h$ with category $c$. The latter states that $t$ is a TARGET of host $h$, i.e., it receives the objects (of category $c$) returned from $h$. By [ShortcutContainer], when both premises hold, we add a shortcut edge from $s$ to $t$.

[HostSource] and [HostTarget] show how to derive $h \overset{c}{\Leftarrow} s$ and $h \overset{c}{\Rightarrow} t$, respectively. The key to derive these two relations is the pointer-host map $pt_H$ as introduced in Section 3.3. The bottom four rules in Figure 10 describe how to compute $pt_H$ for host-related pointers. Specifically, [ColHost] and [MapHost] add a collection or map object $o_i$ ($t_1 <: t_2$ indicates $t_1$ is a subtype of $t_2$) to $pt_H(x)$ if $x$ points to $o_i$. By [TransferHost], if a call site $i$ invokes a TRANSFER method, then we transfer the hosts associated with the receiver variable of $i$ to $i$'s LHS variable, e.g., transfer hosts from $l$ to $it$ for $it = l.iterator()$. By [PropHost], we propagate hosts along PFG edges, except the cases that the edge is a return edge from a TRANSFER method. We exclude such cases because (1) [TransferHost] already handles return values of TRANSFER methods, and (2) different hosts may flow to the same TRANSFER method $m$ and, if we do not exclude such cases, would be propagated to LHS variables of all $m$' s call sites and lead to imprecision.

## 4.4 Local Flow Pattern

Figure 11 gives the rules to compute *cutReturns* and $E_{SC}$ for local flow pattern. To facilitate the formalism, we introduce notation $\rightarrowtail$ as defined by [Param2Var] and [Param2VarRec]. $\langle m, k \rangle \rightarrowtail x$ states that $x$ is a local variable in method $m$, and the objects pointed to by $x$ come from the $k$-th parameter of $m$. Note that [Param2Var] and [Param2VarRec] guarantee that when $\langle m, k \rangle \rightarrowtail x$ holds, the values of $x$ must only come from $m$'s parameter(s) via zero or more local assignments (i.e., the values will not come from other sources such as field loads and method calls). Based on this notation, we could easily define edges to cut off/add for local flow pattern.

*Cut.* In [CutLFlow], if we find that the values of return variable $m_{ret}$ all come from $m$'s parameter(s), then we add $m_{ret}$ to *cutReturns*.

*Shortcut.* In [ShortcutLFlow], if $\langle m, k \rangle \rightarrowtail m_{ret}$ holds, then for each call site $i$ of method $m$, we add a shortcut edge from $k$-th argument of $i$ to its LHS variable.

## 4.5 Soundness

In this section, we give proof sketch for soundness of CUT-SHORTCUT based on the rules we present before. We use $E_{cut}$ to denote the set of edges cut off by our rules. For a TARGET $t$, $Source_t$ represents a set of SOURCES connected to $t$ via shortcut edges, i.e., $Source_t = \{s \mid s \to t \in E_{SC}\}$. We use $PA$ to denote a pointer analysis and add subscript to refer to a specific analysis, e.g., $PA_{ci}$ for context-insensitive pointer analysis and $PA_{csc}$ for CUT-SHORTCUT. For an analysis $PA_a$, $G_a = (N_a, E_a)$ represents its PFG. We show that $PA_{csc}$ can generate sound points-to results for a program.

We first prove that any dynamic pointer flow path from LHS variable of an allocation site (i.e., pointer that first receives a heap object, which is the staring point of the flow) to a pointer $t$ that ends with a cut edge $e$ in $E_{cut}$ must flow through some $s \in Source_t$. It means that, even if we cut off edge $e$, the objects that flow to $t$ via $e$ at run time can still be propagated to $t$ by the shortcut edges to $t$ in $PA_{csc}$. We formally state this in Lemmas 4.1 (for the edges cut by *cutStores*) and 4.2 (for the edges cut by *cutReturns*).

LEMMA 4.1. *A dynamic pointer flow path $P$ that ends with a store edge $y \to o_k.f \in E_{cut}$ must flow through some $s \in Source_{o_k.f}$.*

PROOF SKETCH. Suppose $x.f = y \in cutStores$ and $x$ points to $o_k$. If $P$ ends with $y \to o_k.f$, by [CUTSTORE], it must pass through some arguments passed to $y$. [CUTSTORE] and [PROPSTORE] will search along the call chain from $y$ to find the outermost invocations. $P$ must pass through $y$'s corresponding argument (say $a_y$) at one of those outermost invocations (say $i$). By [SHORTCUTSTORE], $o_k$ must belong to points-to set of $x$'s corresponding argument at $i$, and thus $a_y \in Source_{o_k.f}$.  □

ASSUMPTION 1. *We assume that the input relations ENTRANCES and TRANSFERS for container access pattern are complete w.r.t. the container classes we consider (i.e, container classes in JDK).*

LEMMA 4.2. *Under Assumption 1, a dynamic pointer flow path $P$ that ends with a return edge $m_{ret} \to r \in cutReturns$ must flow through some $s \in Source_r$.*

PROOF SKETCH. We discuss *cutReturns* in each pattern respectively.

- In field access pattern: $m_{ret} \in cutReturns$ is derived from [CUTPROPLOAD]. Suppose $r$ is the LHS variable of invocation $j$. It is easy to see [CUTPROPLOAD] is an induction process. Every element in *tempLoads* arises from an innermost load statement (say $i$). If $P$ flows through either a load edge generated by $i$, since $j_{ak} \mapsto base$, $P$ must flow through $j_{ak}$. By [SHORTCUTLOAD], $j_{ak} \in Source_r$, otherwise, by [RELAYEDGE], other kinds of edges will be connected to $m_{ret}$'s successors.
- In container access pattern: $m_{ret} \in cutReturns$ is derived from [CUTCONTAINER]. By the semantics of EXITS in container access pattern, $r$ should point to all objects added to the related containers (hosts). By Assumption 1, the ENTRANCES and TRANSFERS are complete. By [COLHOST]–[PROPHOST], $pt_H$ will contain all hosts for the related pointers, and by [HOSTSOURCE] as well as [HOSTTARGET], SOURCES and TARGETS will be associated with the hosts. Since $r$ is a TARGET that receives elements from a container, $P$ must flow through some argument(s) passed to the ENTRANCE which will be captured by [HOSTSOURCE] and belong to $Source_r$.
- In local flow pattern: $m_{ret} \in cutReturns$ is derived from [CUTLFLOW]. $\langle m, k \rangle \rightarrowtail m_{ret}$ indicates $m_{ret}$'s points-to set only comes from several parameters of $m$, including $m_{pk}$. Thus, if $P$ ends with $m_{ret} \to r$, it must come from one of these parameters (say $m_{pk}$). Suppose $r$ is the LHS variable of invocation $i$. During dynamic execution, $i_{ak}$ is the predecessor of $m_{pk}$, which means $P$ flows through it. By [SHORTCUTLFLOW], $i_{ak} \in Source_r$.  □

THEOREM 4.3 (SOUNDNESS OF $PA_{csc}$). *Under Assumption 1, $PA_{csc}$ is sound.*

PROOF SKETCH. We will show that for any dynamic pointer flow path $P$ from a starting point (say $s$) to an arbitrary program pointer (say $t$), $t$ is reachable from $s$ on $G_{csc}$. Suppose the nodes in $P$ are $[s = p_0, p_1, ..., p_n = t]$, then we prove $s$ can reach $p_i$ for any $i \in [0, n]$ on $G_{csc}$ by induction.

At first, it holds trivially for $i = 0$, i.e., $s$ reaches itself ($p_0$). For induction, we prove that if $s$ reaches $p_k$ for any $k \le i$ ($i < n$), then it also reaches $p_{i+1}$. Considers the edge $p_i \to p_{i+1}$ on $P$. There are two cases for the edge, (1) $p_i \rightarrow p_{i+1}$ is not cut off on $G_{csc}$, and (2) it is cut off and $G_{csc}$ contains the corresponding shortcut edges for it. In case (1), $s$ can reach $p_{i+1}$ on $G_{csc}$ through path from $s$ to $p_i$ plus $p_i \rightarrow p_{i+1}$. In case (2), by Lemmas 4.1 and 4.2, there exists a $s_{p_{i+1}} \in Source_{p_{i+1}}$ where $s_{p_{i+1}}$ is one of the node in path $[p_0, p_1, ..., p_i]$. Then by the premise of induction, $s$ reaches $s_{p_{i+1}}$, and since there is a shortcut edge $s_{p_{i+1}} \rightarrow p_{i+1}$ on $G_{csc}$, $s$ also reaches $p_{i+1}$. Thus, by induction, we prove $s$ can reach any $p_i$, including $t$, on $G_{csc}$, which means that for every object $o$ that $t$ points to during dynamic execution, $PA_{csc}$ ensures that $o$ is contained in $pt_{csc}(t)$. Hence $PA_{csc}$ is sound. □

## 5 EVALUATION

In this section, we investigate the following research questions for evaluation.

**RQ1.** How does Cut-Shortcut compare to context-insensitive pointer analysis?

**RQ2.** How does Cut-Shortcut compare to conventional context-sensitive pointer analysis?

**RQ3.** How does Cut-Shortcut compare to state-of-the-art selective context-sensitive pointer analysis technique that has similar goal (i.e., for high efficiency with good precision)?

*Implementation.* To thoroughly evaluate the effectiveness of Cut-Shortcut and also demonstrate its generality, we implemented it on two totally diverse pointer analysis frameworks (latest versions), the declarative Doop [DOOP 2022] (written in Datalog) and the imperative Tai-e [Tai-e 2022; Tan and Li 2022] (written in Java). Doop is a pointer analysis framework that has been considered as the mainstream platform to implement and evaluate Java pointer anlaysis in the last decade, and Tai-e is a recent static analysis framework for Java, equipped with a highly efficient pointer analysis system. We fixed an unsound issue of Doop by adding reflection-related classes to the closed world for better soundness. As Datalog prohibits negation in a recursion cycle, we cannot directly implement rule [CutPropLoad] on Doop in Figure 9 (along with rules [Return], [Propagate] and [Call] in Figure 7), and the workaround to get rid of the cycle would increase analysis cost unbearably. Thus, we omit the handling of load in field access pattern for Doop. Overall, Cut-Shortcut's core implementation contains around only 100 Datalog rules on Doop and 2000 lines of Java code on Tai-e. We will release and maintain the source code of Cut-Shortcut on Tai-e at https://github.com/pascal-lab/Tai-e, and the source code on Doop is available in our artifact.

*The Compared Analyses.* We compare Cut-Shortcut to three types of pointer analyses: context insensitivity (CI), mainstream context sensitivity, and state-of-the-art selective context sensitivity. CI is de facto the fastest pointer analysis for Java. For conventional context sensitivity, we select the commonly-used 2-object-sensitive (2obj) [Milanova et al. 2002, 2005] and 2-type-sensitive (2type) [Smaragdakis et al. 2011] analyses. 2obj is highly-precise, and 2type often achieves much better scalability than 2obj while yielding comparable precision. For selective context-sensitivity, we use a state-of-the-art approach Zipper$^e$ [Li et al. 2020a], which exhibits better efficiency and precision trade-off than many other selective approaches. Zipper [Li et al. 2018a] (often evaluated as state-of-the-art in recent literature, with comparable precision but is much less efficient than Zipper$^e$) is not considered as it fails to scale for most of our evaluated programs.

*Experimental Settings.* All experiments were conducted on an Intel Xeon 2.2GHz machine with 128GB of memory. The time budget is set to 2 hours for each analysis. We consider all the large and complex Java programs for evaluation used in recent related literature [Jeon et al. 2019, 2018; Jeong et al. 2017; Li et al. 2018a,b, 2020a; Lu and Xue 2019; Tan et al. 2021] except those that cannot be executed so that we cannot perform recall experiments on them for evaluating soundness. We analyze the programs with a large Java library JDK1.6 that is commonly used in recent work [He et al. 2021, 2022; Jeon et al. 2020; Jeon and Oh 2022; Lu et al. 2021].

*Precision Metrics.* To measure precision, we use four independently useful clients that are widely-used as precision metrics in pointer analysis literature [Jeon et al. 2019, 2020; Jeong et al. 2017; Li et al. 2018a,b, 2020a; Lu et al. 2021; Smaragdakis et al. 2011, 2014; Tan et al. 2017]. The clients are a cast-resolution analysis (metric: the number of casts that may fail—#fail-cast), a method reachability analysis (metric: the number of reachable methods—#reach-mtd), a devirtualization analysis (metric: the number of virtual call sites that cannot be disambiguated into monomorphic calls—#poly-call) and a call-graph building analysis (metric: the number of call graph edges—#call-edge).

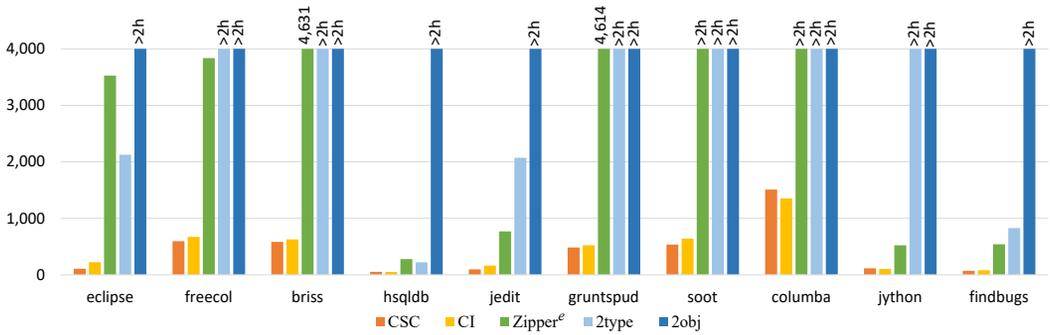

Fig. 12. Analysis time (in seconds) spent by our Cut-Shortcut approach (CSC), context-insensitive (CI), selective 2-object-sensitive (Zipper$^e$) and conventional context-sensitive (2obj, 2type) pointer analyses on Doop. CSC also exhibits similar efficiency advantage over CI on Tai-e (the figure is not shown for limited space) while obtaining significantly better precision than CI on both Doop (Table 1) and Tai-e (Table 2).

## 5.1 RQ1: Cut-Shortcut vs. Context Insensitivity (CI)

In this section, we examine how Cut-Shortcut fares against the fastest pointer analysis, CI. But in order to trust the reliability of Cut-Shortcut's efficiency benefit, we must first confirm its soundness. In addition to the theoretical soundness proofs in Section 4.5, below we further validate its soundness by a recall experiment.

*Soundness (Recall).* In the recall experiment, we execute all the evaluated programs with their default tests (e.g., for DaCapo benchmarks [Blackburn et al. 2006]) or the ones we input (e.g., for GUI programs, we click to interact with them and the results are recorded and will be provided in artifact). Then under these inputs, we dynamically record their reachable methods and call graph edges during execution (it is very hard to instrument other dynamic information like detailed points-to relations), and then examine how many of them can be recalled (over-approximated) by Cut-Shortcut and other analyses in our evaluation. The results show that Cut-Shortcut can recall virtually all true reachable methods and call graph edges discovered by other sound analyses. For the other cases, it misses totally 20 call graph edges for 6 programs on Doop and 11 edges for 3 programs on Tai-e, and after manual inspection, we found that the missed edges are not true and they are incorrectly recorded by the instrumentation tool. As a result, Cut-Shortcut achieves the same soundness as other analyses in the recall experiment. Plus the theoretical soundness proof, we can trust that the remaining efficiency and precision results are reliable.

*Efficiency and Precision.* Figure 12 shows graphically the elapsed time of all analyses on Doop (Tai-e's is not shown for space limitation). Table 1 (for Doop) and Table 2 (for Tai-e) present all efficiency and precision results in full detail.

We can see that CI is substantially more efficient than context-sensitive analyses, and Cut-Shortcut (CSC) is even able to run faster than CI for 7 out of 10 programs on both Doop and Tai-e. For the remaining programs, Cut-Shortcut is either as fast as or negligibly slower than CI. The superb efficiency of Cut-Shortcut stems from its methodology and precision improvements. First, unlike context sensitivity (which improves precision by cloning methods and heap objects under different contexts), Cut-Shortcut improves precision by modifying PFG which generally introduces little overhead. Second, different from existing selective context-sensitivity approaches, PFG does not need a pre-analysis: it performs the context-insensitive pointer analysis algorithm on an on-the-fly constructed PFG. Third, Cut-Shortcut prevents a large amount of spurious points-to results from being propagated to other parts of the program during analysis, and the analysis cost of PFG manipulation is usually negligible compared to the efficiency benefit brought by the increasing precision; as a result, Cut-Shortcut generally exhibits better efficiency than CI.

Table 1. Efficiency and precision results for context-insensitive (CI), conventional context-sensitive (2obj, 2type), and selective 2-object-sensitive (Zipper$^e$) pointer analyses, and our Cut-Shortcut approach (CSC) on the declarative Doop framework. For all numbers, smaller is better.

| Program | Analysis | Time (s) | #fail-cast | #reach-mtd | #poly-call | #call-edge | Program | Analysis | Time (s) | #fail-cast | #reach-mtd | #poly-call | #call-edge |
|---|---|---|---|---|---|---|---|---|---|---|---|---|---|
| eclipse | CI | 223 | 5,077 | 23,549 | 10,744 | 183,478 | gruntspud | CI | 523 | 6,982 | 39,555 | 12,328 | 272,944 |
| | 2obj | >2h | – | – | – | – | | 2obj | >2h | – | – | – | – |
| | 2type | 2,126 | 4,334 | 22,928 | 9,955 | 164,742 | | 2type | >2h | – | – | – | – |
| | Zipper$^e$ | 3,526 | 4,198 | 23,206 | 10,049 | 170,172 | | Zipper$^e$ | 4,614 | 5,846 | 39,092 | 11,277 | 230,831 |
| | CSC | 107 | 4,082 | 23,437 | 10,475 | 173,740 | | CSC | 484 | 5,415 | 39,354 | 11,826 | 251,921 |
| freecol | CI | 671 | 9,971 | 44,838 | 14,626 | 309,033 | soot | CI | 641 | 16,724 | 32,696 | 16,877 | 418,910 |
| | 2obj | >2h | – | – | – | – | | 2obj | >2h | – | – | – | – |
| | 2type | >2h | – | – | – | – | | 2type | >2h | – | – | – | – |
| | Zipper$^e$ | 3,836 | 8,492 | 44,159 | 12,630 | 265,446 | | Zipper$^e$ | >2h | – | – | – | – |
| | CSC | 595 | 7,435 | 44,472 | 13,354 | 287,241 | | CSC | 536 | 10,991 | 32,497 | 16,257 | 399,610 |
| briss | CI | 627 | 8,000 | 41,787 | 12,853 | 294,428 | columba | CI | 1,352 | 10,824 | 56,431 | 17,679 | 413,829 |
| | 2obj | >2h | – | – | – | – | | 2obj | >2h | – | – | – | – |
| | 2type | >2h | – | – | – | – | | 2type | >2h | – | – | – | – |
| | Zipper$^e$ | 4,631 | 6,650 | 41,381 | 11,640 | 261,081 | | Zipper$^e$ | >2h | – | – | – | – |
| | CSC | 582 | 6,129 | 41,563 | 12,130 | 276,029 | | CSC | 1,508 | 8,606 | 56,050 | 17,135 | 397,826 |
| hsqldb | CI | 50 | 1,723 | 11,143 | 1,914 | 61,929 | jython | CI | 104 | 2,381 | 12,623 | 2,937 | 119,227 |
| | 2obj | >2h | – | – | – | – | | 2obj | >2h | – | – | – | – |
| | 2type | 221 | 1,276 | 10,769 | 1,598 | 54,652 | | 2type | >2h | – | – | – | – |
| | Zipper$^e$ | 278 | 1,209 | 10,793 | 1,618 | 54,725 | | Zipper$^e$ | 523 | 2,193 | 12,423 | 2,796 | 115,639 |
| | CSC | 52 | 1,295 | 10,986 | 1,744 | 56,513 | | CSC | 115 | 1,886 | 12,325 | 2,904 | 111,159 |
| jedit | CI | 163 | 4,349 | 25,157 | 6,336 | 149,761 | findbugs | CI | 81 | 3,488 | 16,985 | 4,575 | 106,233 |
| | 2obj | >2h | – | – | – | – | | 2obj | >2h | – | – | – | – |
| | 2type | 2,071 | 3,458 | 24,475 | 5,493 | 124,798 | | 2type | 827 | 2,472 | 16,512 | 3,866 | 88,923 |
| | Zipper$^e$ | 768 | 3,401 | 24,552 | 5,594 | 125,883 | | Zipper$^e$ | 541 | 2,495 | 16,696 | 4,038 | 93,485 |
| | CSC | 97 | 3,188 | 24,629 | 5,880 | 129,671 | | CSC | 70 | 2,568 | 16,782 | 4,287 | 93,800 |

Now let us examine the precision. For all precision metrics, the smaller number signifies a better result. Cut-Shortcut achieves better precision than CI for all precision metrics on both Doop and Tai-e, and the precision improvement is usually significant, especially for #fail-cast and #call-edge. Cut-Shortcut affects precision via modifying PFG guided by the patterns, and the improvements show that the three patterns we designed for Cut-Shortcut are useful for improving precision.

One may wonder which pattern of Cut-Shortcut has the biggest impact on precision? To address this question, we conduct three sets of experiments for Cut-Shortcut on Tai-e to measure the impact of the three patterns for different clients. For each experiment, only one pattern is enabled and the other two patterns are disabled. To measure the impact, we compute the precision improvement over CI of each individual pattern and divide it by the overall improvement of three patterns together. We found the impact of patterns varies for different clients. For example, on average, field access pattern, container pattern, and local flow pattern improve the precision by 11.9%, 75.8% and 11.8% respectively for client #fail-cast, and 53.2%, 40.5% and 2.0% for client #reach-mtd. (Note that the sum of the three percentage numbers is less than 100%, because their combination can further help improve the precision of each one by reducing the spurious value propagation during the analysis).

Table 2. Efficiency and precision results for context-insensitive (CI), conventional context-sensitive (2obj, 2type), and selective 2-object-sensitive (ZIPPER$^e$) pointer analyses, and our CUT-SHORTCUT approach (CSC) on the imperative TAI-E framework. For all numbers, smaller is better.

| Program | Analysis | Time (s) | #fail-cast | #reach-mtd | #poly-call | #call-edge | Program | Analysis | Time (s) | #fail-cast | #reach-mtd | #poly-call | #call-edge |
|---|---|---|---|---|---|---|---|---|---|---|---|---|---|
| eclipse | CI | 21 | 5,091 | 23,920 | 10,610 | 184,294 | gruntspud | CI | 44 | 6,775 | 39,800 | 12,342 | 274,872 |
| | 2obj | 5,218 | 3,659 | 23,195 | 9,690 | 164,143 | | 2obj | >2h | – | – | – | – |
| | 2type | 261 | 4,230 | 23,338 | 9,876 | 165,823 | | 2type | >2h | – | – | – | – |
| | ZIPPER$^e$ | 557 | 4,154 | 23,583 | 9,907 | 170,947 | | ZIPPER$^e$ | 162 | 5,308 | 39,201 | 10,978 | 232,046 |
| | CSC | 12 | 3,977 | 23,746 | 10,198 | 173,556 | | CSC | 39 | 5,035 | 39,444 | 11,675 | 240,281 |
| freecol | CI | 54 | 9,779 | 46,896 | 15,496 | 323,141 | soot | CI | 107 | 16,659 | 32,918 | 16,372 | 415,728 |
| | 2obj | >2h | – | – | – | – | | 2obj | >2h | – | – | – | – |
| | 2type | >2h | – | – | – | – | | 2type | >2h | – | – | – | – |
| | ZIPPER$^e$ | 179 | 7,199 | 46,064 | 13,546 | 279,502 | | ZIPPER$^e$ | 1,845 | 12,901 | 32,178 | 14,953 | 281,751 |
| | CSC | 40 | 6,511 | 46,261 | 14,143 | 288,839 | | CSC | 51 | 10,509 | 32,660 | 15,152 | 337,562 |
| briss | CI | 50 | 7,815 | 41,851 | 12,755 | 294,038 | columba | CI | 117 | 10,492 | 56,787 | 17,700 | 425,149 |
| | 2obj | >2h | – | – | – | – | | 2obj | >2h | – | – | – | – |
| | 2type | >2h | – | – | – | – | | 2type | >2h | – | – | – | – |
| | ZIPPER$^e$ | 179 | 6,098 | 41,427 | 11,575 | 261,811 | | ZIPPER$^e$ | 1,163 | 8,257 | 55,973 | 15,931 | 346,346 |
| | CSC | 58 | 5,669 | 41,462 | 11,891 | 265,948 | | CSC | 122 | 7,862 | 56,427 | 16,963 | 389,537 |
| hsqldb | CI | 4 | 1,674 | 11,588 | 1,817 | 64,393 | jython | CI | 11 | 2,421 | 13,058 | 2,951 | 121,528 |
| | 2obj | >2h | – | – | – | – | | 2obj | >2h | – | – | – | – |
| | 2type | 25 | 1,094 | 11,185 | 1,472 | 56,712 | | 2type | >2h | – | – | – | – |
| | ZIPPER$^e$ | 26 | 1,027 | 11,214 | 1,496 | 56,827 | | ZIPPER$^e$ | 42 | 1,898 | 12,617 | 2,597 | 111,866 |
| | CSC | 3 | 1,169 | 11,335 | 1,618 | 57,986 | | CSC | 11 | 1,864 | 12,683 | 2,884 | 112,364 |
| jedit | CI | 15 | 4,113 | 25,261 | 6,339 | 150,159 | findbugs | CI | 6 | 3,391 | 17,353 | 4,462 | 107,339 |
| | 2obj | 7,192 | 2,578 | 24,214 | 4,933 | 121,896 | | 2obj | 1,369 | 1,984 | 16,821 | 3,560 | 88,890 |
| | 2type | 407 | 3,104 | 24,263 | 5,097 | 122,782 | | 2type | 83 | 2,379 | 16,874 | 3,752 | 90,018 |
| | ZIPPER$^e$ | 39 | 3,042 | 24,342 | 5,212 | 123,783 | | ZIPPER$^e$ | 22 | 2,453 | 17,099 | 3,917 | 95,543 |
| | CSC | 10 | 2,913 | 24,641 | 5,825 | 129,248 | | CSC | 5 | 2,215 | 17,099 | 4,167 | 92,451 |

## 5.2 RQ2: CUT-SHORTCUT vs. Conventional Context Sensitivity

In this section, we investigate how CUT-SHORTCUT performs compared to mainstream conventional context sensitive pointer analyses, 2obj and 2type.

*Efficiency.* As shown in Tables 1 and 2, 2obj runs out of time for all 10 programs on DOOP and 7 out of 10 programs on TAI-E. 2type is more scalable than 2obj, but it still fails to scale for 6 programs on both DOOP and TAI-E. As a comparison, CUT-SHORTCUT can finish analysis for all programs. For the programs that 2obj and 2type scale for, CUT-SHORTCUT runs remarkably faster. For 2obj, CUT-SHORTCUT is on average 475.9× faster on TAI-E (2obj is not scalable for all programs on DOOP, so we cannot compute its speedup), and for 2type, the average speedup of CUT-SHORTCUT is 14.3× on DOOP and 21.8× on TAI-E. In summary, CUT-SHORTCUT achieves dramatically better efficiency and higher scalability than conventional context sensitivity.

*Precision.* When 2obj is scalable (for 3 out of 10 programs on TAI-E), it is more precise than all other analyses including CUT-SHORTCUT for all precision metrics. When 2type is scalable (for 4 out of 10 programs on DOOP and TAI-E), it is generally more precise than CUT-SHORTCUT, but it performs worse on #fail-cast for 2 out of 4 programs on DOOP and 3 out of 4 programs on TAI-E. These results demonstrate the high precision of conventional context-sensitivity approaches; however, we cannot rely on them to improve precision in many programs due to their poor scalability.

## 5.3 RQ3: CUT-SHORTCUT vs. State-of-the-Art Selective Context Sensitivity

In this section, we compare CUT-SHORTCUT to ZIPPER$^e$ [Li et al. 2020a], a state-of-the-art selective context sensitivity approach. We use the default configuration of ZIPPER$^e$ as in [Li et al. 2020a], which is well-tuned and achieves very good trade-off between efficiency and precision.

Table 3. Detailed comparison between Zipper$^e$ and Cut-Shortcut on Doop (left half) and Tai-e (right half).

| Program | Zipper$^e$ | | | | CSC | | | Zipper$^e$ | | | | CSC | | |
|---|---|---|---|---|---|---|---|---|---|---|---|---|---|---|
| | Total time | Pre-analysis | Main analysis | Selected methods | Time | Involved methods | Overlapped methods | Total time | Pre-analysis | Main analysis | Selected methods | Time | Involved methods | Overlapped methods |
| eclipse | 3,526 | 251 | 3,275 | 3,616 | 107 | 3,827 | 42.8% | 557 | 29 | 528 | 3,430 | 12 | 4,168 | 34.3% |
| freecol | 3,836 | 992 | 2,844 | 4,497 | 595 | 7,650 | 27.7% | 179 | 109 | 70 | 3,855 | 40 | 7,211 | 22.7% |
| briss | 4,631 | 974 | 3,657 | 3,931 | 582 | 7,208 | 24.3% | 179 | 100 | 79 | 3,206 | 58 | 6,990 | 19.0% |
| hsqldb | 278 | 57 | 221 | 1,846 | 52 | 1,914 | 41.0% | 26 | 7 | 19 | 1,494 | 3 | 1,819 | 32.8% |
| jedit | 768 | 214 | 554 | 2,811 | 97 | 5,241 | 25.2% | 39 | 28 | 11 | 2,245 | 10 | 4,241 | 22.3% |
| gruntspud | 4,614 | 1,066 | 3,548 | 4,058 | 484 | 6,980 | 25.8% | 162 | 101 | 61 | 3,432 | 39 | 6,506 | 22.3% |
| soot | >2h | 806 | >2h | 9,802 | 536 | 4,585 | 60.6% | 1,845 | 121 | 1,724 | 12,436 | 51 | 6,072 | 67.2% |
| columba | >2h | 1,936 | >2h | 6,137 | 1,508 | 9,322 | 27.0% | 1,163 | 352 | 811 | 4,247 | 122 | 9,072 | 19.7% |
| jython | 523 | 324 | 199 | 786 | 115 | 2,504 | 14.9% | 42 | 33 | 9 | 1,534 | 11 | 2,557 | 20.3% |
| findbugs | 541 | 93 | 448 | 2,438 | 70 | 3,037 | 36.4% | 22 | 11 | 11 | 2,188 | 5 | 2,909 | 33.0% |

*Efficiency.* The execution of Zipper$^e$ consists of three parts: pre-analysis (a context-insensitive pointer analysis), Zipper$^e$ itself (to select a set of methods according to its strategy), and main analysis (the pointer analysis that applies context sensitivity only to the selected methods); hence, the elapsed time of Zipper$^e$ is the sum of execution time of the three parts. Tables 1 and 2 clearly show that Zipper$^e$ is more scalable and faster than 2obj and 2type on both Doop and Tai-e. On Doop, Zipper$^e$ scales for 8 out of 10 programs, and on Tai-e, Zipper$^e$ scales for all programs. Nonetheless, Cut-Shortcut is still superior to Zipper$^e$ in terms of efficiency with an average speedup of 10.3× on Doop and 12.5× on Tai-e for the programs that Zipper$^e$ can scale. To thoroughly compare the efficiency of Zipper$^e$ and Tai-e, we give detailed elapsed time of Zipper$^e$ and Cut-Shortcut in Table 3, where the left half and right half of the table show data on Doop and Tai-e, respectively. For Zipper$^e$, column "Total time" lists total analysis time of the three parts of Zipper$^e$ (same as in Figure 1 and 2), column "Pre-analysis" shows the summation of analysis time for CI and Zipper$^e$ itself, and column "Main analysis" gives analysis time for Zipper$^e$-guided context-sensitive pointer analysis. As we can see, the cost of pre-analysis of Zipper$^e$ occupies 25.6% (on Doop) and 44.8% (on Tai-e) of Zipper$^e$'s total time on average, and even if we only consider the main analysis of Zipper$^e$, Cut-Shortcut is still 8.4× (on Doop) and 10× (on Tai-e) faster than Zipper$^e$, which again demonstrates Cut-Shortcut's high efficiency.

*Precision.* For #fail-cast, Cut-Shortcut outperforms Zipper$^e$ for all programs except findbugs on Doop and hsqldb on both. For the remaining precision metrics, Cut-Shortcut is still comparable to Zipper$^e$ (despite the high advantage of analysis speed of Cut-Shortcut). For some programs like jython on Doop, Cut-Shortcut even outperforms Zipper$^e$ in 3 out 4 precision metrics. We expect that by conducting additional program patterns based on Cut-Shortcut's principle in the future, we can further improve the precision (and efficiency) of Cut-Shortcut.

Although Zipper$^e$ and Cut-Shortcut improve precision in fundamentally different ways, one may wonder whether the methods selected by Zipper$^e$ are the same as the ones that are involved in the cut and shortcut edges identified by Cut-Shortcut? Table 3 answers this question in detail. On average (including both Doop and Tai-e), Cut-Shortcut considers 5191 methods (that are involved in the cut and shortcut edges), accounting for 17% of all reachable methods per program, while Zipper$^e$ selects 3899 methods, and only 31% of methods involved in the edges identified by Cut-Shortcut are also selected by Zipper$^e$ (see column "Overlapped methods").

## 6 RELATED WORK

Context sensitivity plays an essential role for whole-program Java pointer analysis, and we have discussed some of the related work in earlier sections. The other relevant research is covered below.

In recent years, many selective context-sensitivity approaches are proposed to make efficiency and precision trade-off, especially for large and complex Java programs. We can understand "selective" in two ways: select more effective context elements to analyze each method (*Select context elements*) and select a set of program methods that need context sensitivity to analyze them (*Select program methods*). Below we discuss these two types of selective context-sensitivity approaches in turn.

*Select Context Elements.* Conventional ($k$-limiting) context sensitivity uses $k$ consecutive context elements, e.g., $[c_k..., c_2, c_1]$ to analyze a method $m$, where $c_1$ is $m$'s call site and $c_2$ is the call site of the method containing $c_1$, etc. However, Tan et al. [2016] found that this approach may result in many context elements that are not useful for improving precision while occupying the slots of context elements limited by $k$. Therefore, Tan et al. [2016] present an approach to recognize such redundant context elements by exploiting the so-called object allocation graph they proposed. Similarly, Jeon et al. [2018] develop a machine-learning scheme called context tunneling to select such precision-useless context elements for more effective analysis. Further, on the basis of context tunneling, [Jeon and Oh 2022] propose an interesting approach that transforms object sensitivity to call-site sensitivity and show that the latter can simulate the former (but not vice versa).

*Select Program Methods.* Conventional context sensitivity uniformly applies contexts to every method in a given program. However, for certain program methods, context sensitivity does not help improve precision but only incurs extra analysis cost. As a result, researchers propose to select a set of methods that are beneficial to analysis precision and only apply context sensitivity to them while analyzing the remaining methods context-insensitively. To make a good efficiency and precision trade-off, the overall principle is to select the methods that are precision-critical [Li et al. 2018a] but not scalability-threaten [Li et al. 2020a]. To do so, various selection strategies exist: they rely on parameterized heuristics (whose thresholds are based on expert experience) [Hassanshahi et al. 2017; Smaragdakis et al. 2014], machine-learning approaches [Jeon et al. 2019, 2020; Jeong et al. 2017], abstracted memory capacity [Li et al. 2018b], program patterns [Li et al. 2018a, 2020a] or the scheme that is able to take advantages of the above ones [Tan et al. 2021]; in addition, context sensitivity can also be applied to the selected variables (rather than methods) [He et al. 2021; Lu and Xue 2019] based on CFL-reachability [Reps 1998].

*Hybrid Context Sensitivity.* In some pointer analyses, the context elements for the same method may vary. Kastrinis and Smaragdakis [2013] present a hybrid context-sensitivity approach in which a method may be analyzed under both call-site and object sensitivity, and in some cases, such combination may lead to more effective results than the single type of context elements. Thakur and Nandivada [2020] mix object-sensitivity contexts and the level-summarized relevant value contexts (a context abstraction proposed in their earlier work [Thakur and Nandivada 2019]) to analyze each method for yielding more precise results than individual of them.

No matter the selective or the hybrid context-sensitivity approaches described above, they still rely on the core idea of context sensitivity to replicate (a set of) methods and analyze the program elements in them separately under different (types of) contexts. Our approach is fundamentally different in that it simulates the effect of context sensitivity by cutting off precision-loss edges and add shortcuts directly from Source to Target on PFG as explained throughout the paper.

*Other Related Work.* In terms of graph modification, Li et al. [2020b, 2022] introduce an algorithm to simplify the input graph of interleaved Dyck reachability framework which is able to express various analysis problems (e.g., taint analysis [Huang et al. 2015], demand-driven context-sensitive pointer analysis [Späth et al. 2016; Sridharan and Bodík 2006]). Cut-Shortcut differs from it in both the target problem and the underlying graph. Speaking of removing spurious points-to information for Java, in addition to context sensitivity, De and D'Souza [2012] present a flow-sensitive approach to partially perform strong updates to heap objects. They accomplish this by expressing points-to relations via access paths [Fink et al. 2008; Kanvar and Khedker 2016]; differently, Cut-Shortcut and many other schemes often express points-to information via PFG-like graphs. Trace partitioning [Blanchet et al. 2003; Rival and Mauborgne 2007] analyzes the code `if (c) {S1} else {S2} <rest>` as `if (c) {S1;<rest>} else {S2;<rest>}`. The `<rest>` is cloned twice into both branches and then the merging at the original starting point of `<rest>` is delayed to the end of the `<rest>`. This is more similar to the idea of context sensitivity which clones code. Cut-Shortcut is fundamentally different as we do not clone code; moreover, instead of delaying merging, we avoid merging by removing the edges along which imprecision propagates outside a method for improving precision. Zhang et al. [2014] present a hybrid top-down and bottom-up analysis. Top-down and bottom-up approaches are general ideas that many static analyses adopt. Like other mainstream pointer analyses for Java [Smaragdakis and Balatsouras 2015], Cut-Shortcut also performs in a top-down manner; if we treat bottom-up analysis as a summarization, the high-level idea of our addition of shortcut edges could be seen as a kind of pointer-specific summarization. However, neither the target problem, nor its underlying concrete methodology of Cut-Shortcut is similar to [Zhang et al. 2014]. Finally, there are efforts to accelerate context-sensitive Java pointer analysis by merging heap objects [Chen et al. 2021; Tan et al. 2017], which are orthogonal to Cut-Shortcut.

## 7 CONCLUSIONS

For the past 20 years, context sensitivity has been considered as virtually the most useful technique for increasing the precision of Java pointer analysis. However, it brings heavy efficiency costs especially for large and complex programs. Selective context-sensitivity approaches partially alleviate this issue but have not solved its efficiency bottleneck, because it is very hard to correctly select the methods that are precision-critical but do not threaten scalability; as a result, they still run the risk of computing and maintaining a large number of contexts to distinguish spurious object flows, which finally yields limited results.

To address this dilemma, we present a fundamentally different approach called Cut-Shortcut which tries to simulate the effect of context sensitivity without applying contexts. This is achieved by cutting off the edges that bring precision loss and add shortcut edges for connecting related pointers on the typical PFG of pointer analysis. We instantiate Cut-Shortcut by exploiting three program patterns and designing rules based on them in accordance with Cut-Shortcut's principle. Then we formalize it and prove its soundness. To comprehensively validate Cut-Shortcut's effectiveness (and generality as a principled approach), we implement two versions of it for two state-of-the-art Java pointer analysis frameworks: one in Datalog for the declarative Doop and the other in Java for the imperative Tai-e. The experimental results are extremely promising: Cut-Shortcut is even able to run faster than context insensitivity for most evaluated programs while obtaining significantly better precision (high precision that is comparable to context sensitivity) in both frameworks. Given the encouraging outcomes, more program patterns or insights are expected to be explored on top of Cut-Shortcut, and we hope that our approach could provide some new perspectives for developing more effective pointer analysis for Java in the future.

## ACKNOWLEDGMENTS

We thank our shepherd Alexey Loginov and anonymous reviewers for their helpful comments. This work was supported by the Natural Science Foundation of China under Grant Nos. 62025202, 62002157 and 61932021, and the Leading-edge Technology Program of Jiangsu Natural Science Foundation under Grant No. BK20202001. The authors would also like to thank the support from the Collaborative Innovation Center of Novel Software Technology and Industrialization, Jiangsu, China.

## DATA AVAILABILITY STATEMENT

We have provided an artifact [Ma et al. 2023] to reproduce all the tables and figures in Section 5, as well as the recall experiment in Section 5.1. You can download the artifact from https://doi.org/10.5281/zenodo.7808384, and use it to reproduce the results by following the instructions given in README.pdf in the artifact package.